\documentclass[aip,reprint,amsmath,amssymb]{revtex4-1}

\usepackage{graphicx}
\usepackage{bm}
\usepackage{siunitx}
\usepackage{subcaption}
\usepackage[none]{hyphenat}
\usepackage{ragged2e}
\usepackage{tabularx}
\usepackage{hhline}
\usepackage[inline]{enumitem}
\usepackage{sidecap}
\usepackage{afterpage}
\usepackage{color}
\usepackage{hyperref}
\usepackage{cleveref}

\DeclareCaptionJustification{justified}{\justifying}
\newcommand{\subfigLabel}[1]{\begin{subfigure}{0pt}
		\phantomcaption
		\label{#1}
	\end{subfigure}}

\Crefname{figure}{Fig.}{Figs.}
\crefname{equation}{Eq.}{Eqs.}
\crefname{table}{Table}{Tables}

\begin{document}

\title{W-waveform Standing Surface Acoustic Waves with Two Equilibrium Positions under Linear Phase Modulation for Patterning Microparticles into Alternate Grid Patterns}

\author{Junseok Lee}
\email{junseok\_lee@yonsei.ac.kr}
\affiliation{School of Mechanical Engineering, Yonsei University, Seoul, Korea}

\date{\today}

\begin{abstract}
This paper presents W-waveform Standing Surface Acoustic Waves (W-SSAW), and as its application, patterning of two groups of microparticles with different sizes alternately without fixing firstly patterned particles.
W-SSAW is constructed by two standing surface acoustic waves of frequencies $f$ and $2f$.
Combined with linear phase modulation to translate Gor'kov potential at a constant speed, W-SSAW can selectively trap particles.
The trapped particles follow the moving Gor'kov potential maintaining force equilibrium between Stokes' drag and the radiation force by W-SSAW.
There exist two asymmetric equilibrium positions every period, and by the asymmetry, each group of particles is trapped at different equilibrium positions to form an alternate pattern.
This technique is extended to two-dimensional alternate patterning by maintaining phase difference \SI{90}{\degree} between X- and Y-directional W-SSAWs.
The patterning method utilizing W-SSAW is advantageous over SSAW-based patterning in that it does not require temporal separation to fix firstly patterned particles.
\end{abstract}

\maketitle

\begin{figure*}[!t]
	\begin{minipage}[l]{0.667\textwidth}%
		\includegraphics[width=\textwidth]{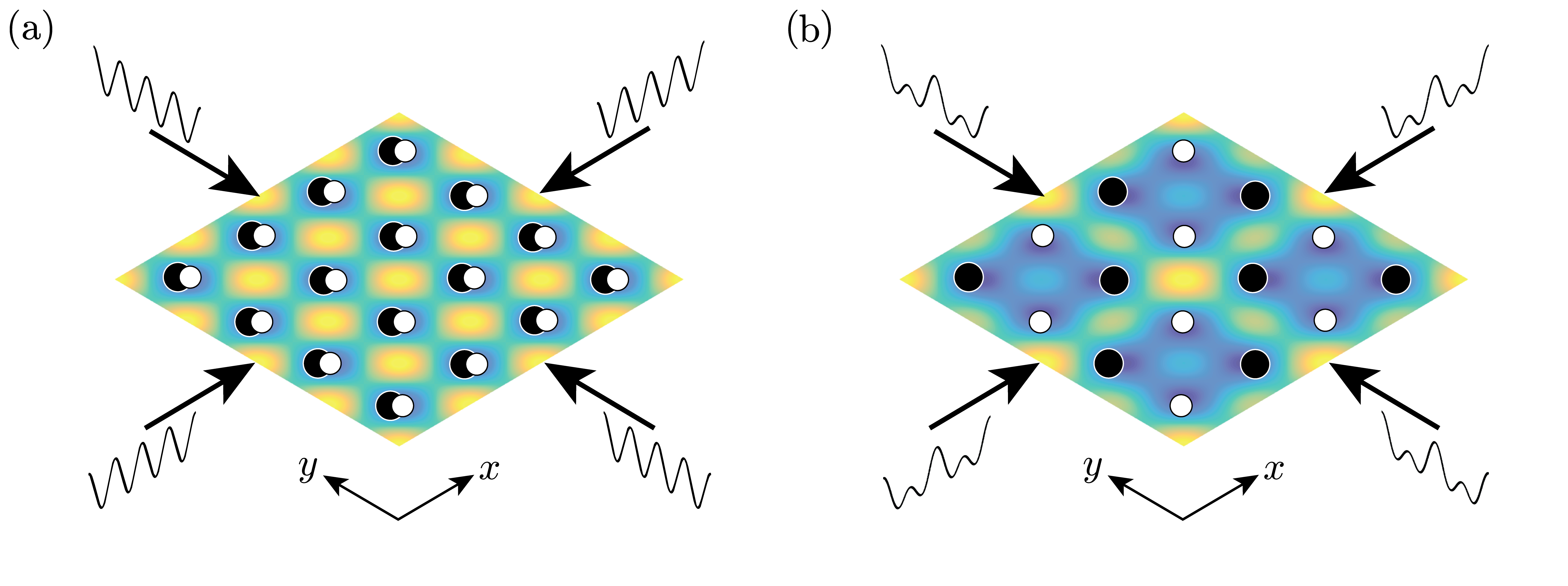}%
	\end{minipage}\hfill%
	\subfigLabel{fig:conv-SSAW}%
	\subfigLabel{fig:W-SSAW}%
	\begin{minipage}[c]{0.3\textwidth}%
		\caption{
			\textbf{Comparison of patterning results using SSAW and W-SSAW, depicted on Gor'kov potentials.}
			Black and white spheres represent two groups of particles which have different sizes, but have the same sign of constrast factors.
			\subref{fig:conv-SSAW} Patterning utilizing SSAW cannot separate each group of particles into different positions, as Gor'kov potential of SSAW only has one equilibrium position every period.
			\subref{fig:W-SSAW} In W-SSAW, there exist two asymmetric equilibrium positions every period. They allow each group of particles to be patterned into different equillibrium positions forming an alternate grid pattern.
			}
	\end{minipage}
\end{figure*}
Recent scientific advances have enabled building substrates with desired patterns to accurately arrange micro- and nano-objects including living cells.
Patterning cells and microparticles into desired arrangements is an important step in biological studies and applications such as tissue engineering, differentiation-on-a-chip, and cellar microenvironment manipulation.\cite{stevens2005direct,tourovskaia2005differentiation,goubko2009patterning}
In addition to passive methods that arrange micro-objects using prefabricated patterns, several active patterning methods that incorporate optics, electromagnetics, and acoustics have been demonstrated.\cite{grier2003revolution,ino2008cell,lee2004manipulation,lee2007integrated,albrecht2006probing,gray2004dielectrophoretic,mittal2007ndep,rosenthal2005dielectrophoretic,shi2009acoustic}
Especially, patterning with acoustic tweezers has been spotlighted due to the nature of acoustic waves which permits patterning of a mass of cells simultaneously without damaging or labeling cells.\cite{ding2013surface}

Surface acoustic wave (SAW) is an elastic wave traveling along the surface storing most of its energy within the surface.
Two SAWs propagating in the opposite direction form standing SAW (SSAW), which exerts an acoustic radiation force on suspended particles as a result of scattering of the waves.
The acoustic radiation force by SSAW moves particles with positive contrast factors to pressure nodes, and conversely, particles with negative ones to antinodes.
Using SSAW, two-dimensional patterning of identical particles into the same pressure nodes is achieved.\cite{shi2009acoustic}
In addition, patterning microparticles with positive contrast factors into pressure nodes, and particles with negative ones into antinodes to form alternate grid patterns is also accomplished.\cite{owens2016}
However, when two groups of particles have the same sign of contrast factors, SSAW only arranges the particles at the same nodes or antinodes to form simple grid patterns [\Cref{fig:conv-SSAW}].
Thus far, arranging particles with the same sign of contrast factor into an alternate grid pattern using SSAW has been based on temporal separation to fix the firstly patterned particles.\cite{gesellchen2014cell}
Only after one group of particles is firstly patterned and fixed on a substrate, the other group is patterned between them.

This paper presents a method of patterning particles alternately using W-waveform SSAW (W-SSAW).
W-SSAW is constructed by two SSAWs that are of frequencies $f$ and $2f$.
It is shown that the radiation force exerted by W-SSAW on a particle  equals linear addition of acoustic radiation forces by two SSAWs, and makes Gor'kov potential W-shaped to have asymmetry.
On the other hand, when linear phase modulation is applied to W-SSAW to translate Gor'kov potential with a uniform speed, only particles that can maintain force equilibrium between Stokes' drag and the radiation force by W-SSAW are trapped by moving Gor'kov potential.
From the asymmetry in Gor'kov potential, there emerge two asymmetric equilibrium positions, and each group of particles is trapped at different equilibrium positions forming an alternate grid pattern [\Cref{fig:W-SSAW}].

Two major forces in acoustofluidics are acoustic radiation force and Stokes' drag force. Acoustic radiation force is given as
\begin{subequations}
\begin{gather}
\mathbf{F}_r  =  F_0 \sin \left( {2kx} \right) \mathbf{ \hat e}_x ,\label{eqn:ARF} \\
F_0  = \frac{{\pi p_0^2{V_c}{\beta _w}}}{{2\lambda }}\varphi,  \label{eqn:ARF0}
\end{gather}
\end{subequations}
where $k$, $p_0$, $V_c$, $\lambda$, and $\varphi$ are wavenumber, acoustic pressure,  the volume of the particle, wavelength, and contrast factor respectively.\cite{collins2015two}
The contrast factor is a material property, which is calculated as
\begin{equation}
	\varphi  = \frac{{5{\rho _c} - 2{\rho _w}}}{{2{\rho _c} + {\rho _w}}} - \frac{{{\beta _c}}}{{{\beta _w}}},
\end{equation}
where  $\rho_c$, $\rho_w$, $\beta_c$ and $\beta_w$ are the density of the particle, the density of a fluid, the compressibility of the particle and the compressibility of the fluid.
When phase modulation $\phi$ is applied, pressure nodes and antinodes are displaced by $\phi/2k$. Hence, acoustic radiation force can be obtained by substituting $x \to x - \phi /2k$ in \cref{eqn:ARF}, giving
\begin{equation} \label{eqn:modARF}
\mathbf{F}_r  =  F_0 \sin \left( {2kx-\phi} \right) \mathbf{ \hat e}_x.
\end{equation}

The Stokes' drag in a quiescent fluid is given as
\begin{subequations}
\begin{align} 
\mathbf{F}_d & =  - b \mathbf{u}, \label{eqn:stokes} \\
		   b & = 6\pi a\eta,
\end{align}
\end{subequations}
where $\mathbf{u}$, and $\eta$ are the particle velocity, and fluid viscosity, respectively.\cite{bruus11}

Acoustic radiation force is derived from ${\mathbf F}_r=-\nabla U$, where $U$ is Gor'kov acoustic potential.
The Gor'kov potential is given by
\begin{equation} \label{eqn:gorkov}
U = V_c \left[ {{\xi_1}\frac{1}{2}{\kappa _0}\left\langle {p_{{\rm{in}}}^2} \right\rangle  - {\xi_2}\frac{3}{4}{\rho _0}\left\langle {\mathbf{v}_{{\rm{in}}}^2} \right\rangle } \right],
\end{equation}
where $V_c$, $\rho_0$, $\kappa_0$, ${p_{{\rm{in}}}}$, ${\mathbf{v}_{{\rm{in}}}}$, $\xi_1$ and $\xi_2$ are the volume of a particle, the density of a fluid, the compressibility of a fluid, the pressure field of incident wave, the velocity vector field of the wave, the monopole coefficient and the dipole coefficient, respectively.\cite{bruus12}
The angle brackets represent time average operator.

Acoustic pressure field $p_{ij}$ and velocity vector field $\mathbf{v}_{ij}$ of $x_i$-directional SSAW of frequency $f_j$ are expressed as
\begin{subequations}
\begin{align}
	p_{ij}\left(x_i,t\right) & = A_{ij}\cos \left(k_{ij}x_i\right)\sin \left( \omega_{ij} t + \phi_{ij} \right), \\
	{\mathbf{v}}_{ij}\left(x_i,t\right) & = v_{ij}\sin \left(k_{ij}x_i\right)\cos \left( \omega_{ij} t + \phi_{ij} \right) \mathbf{\hat e}_i,
\end{align}
\end{subequations}
where $A_{ij}$, $v_{ij}$, $k_{ij}$, $\omega_{ij}$, and $\phi_{ij}$ denote pressure amplitude, velocity amplitude, wave number, angular velocity and phase.
From the superposition principle, the pressure fields and velocity fields of W-SSAW are obtained as
\begin{subequations} \label{eqn:pv1w}
\begin{align}
p_{1 {\rm w}} & = p_{11}\left(x,t\right) + p_{12}\left(x,t\right), \\
\mathbf{v}_{1 \rm{w}} & =  \mathbf{v}_{11}\left(x,t\right)+\mathbf{v}_{12}\left(x,t\right),
\end{align}
\end{subequations}
where $p_{1 {\rm w}}$ and $\mathbf{v}_{1 \rm{w}}$ denote pressure field and velocity vector field of X-directional W-SSAW.
The root mean square for $p_{1 \rm{w}}$ is expressed as
\begin{widetext}
	\begin{align}
\left\langle {p_{1{\rm{w}}}^2} \right\rangle & = {\left\langle {A_{11}^2{{\cos }^2}\left( {{k_{11}}x} \right){{\sin }^2}\left( {{\omega _{11}}t + {\phi _{11}}} \right)} \right\rangle  + \left\langle {A_{12}^2{{\cos }^2}\left( {{k_{12}}x} \right){{\sin }^2}\left( {{\omega _{12}}t + {\phi _{12}}} \right)} \right\rangle } \nonumber \\
&\quad { + \left\langle {2{A_{11}}{A_{12}}\cos \left( {{k_{11}}x} \right)\cos \left( {{k_{12}}x} \right)\sin \left( {{\omega _{11}}t + {\phi _{11}}} \right)\sin \left( {{\omega _{12}}t + {\phi _{12}}} \right)} \right\rangle } \nonumber \\
& = {\left\langle {p_{11}^2} \right\rangle  + \left\langle {p_{12}^2} \right\rangle  + 2{A_{11}}{A_{12}}\cos \left( {{k_{11}}x} \right)\cos \left( {{k_{12}}x} \right)\left\langle {\sin \left( {{\omega _{11}}t + {\phi _{11}}} \right)\sin \left( {{\omega _{12}}t + {\phi _{12}}} \right)} \right\rangle }. \label{eqn:rmsp1w}
	\end{align}
\end{widetext}
Considering W-SSAW is comprised of two SSAWs of frequencies $f$ and $2f$, from the orthogonality of trigonometric functions, the last term in \cref{eqn:rmsp1w} vanishes, giving
\begin{equation} \label{eqn:pfield}
\left\langle {p_{1{\rm{w}}}^2} \right\rangle  = {\rm{ }}\left\langle {p_{11}^2} \right\rangle  + \left\langle {p_{12}^2} \right\rangle.
\end{equation} 
It should be noted that \cref{eqn:pfield} holds true regardless of phase $\phi_{11}$ and $\phi_{12}$.
Similarly, the root mean square for ${\mathbf{v}_{{\rm{in}}}}$ is expressed as
\begin{equation} \label{eqn:vfield}
\left\langle {\mathbf{v}_{{1\rm{w}}}^2} \right\rangle  = \left\langle {\mathbf{v}_{11}^2} \right\rangle  + \left\langle {\mathbf{v}_{12}^2} \right\rangle.
\end{equation}
By substituting \cref{eqn:pfield,eqn:vfield} into \cref{eqn:gorkov}, we obtain
\begin{equation}
{U_{1 \rm{w}}} = U_{11} + U_{12},
\end{equation}
where $U_{11}$ and $U_{12}$ are Gor'kov potentials of SSAWs of frequencies $f$ and $2f$, respectively.
Therefore, acoustic radiation force of W-SSAW is expressed as
\begin{align}
{{\bf{F}}_{1 \rm{w}}} & = - \nabla {U_{1 \rm w}} = {\bf{F}}_{11} + {\bf{F}}_{12} \nonumber \\
	& = F_{01}\sin\left(2kx\right) \mathbf{ \hat e}_x+ F_{02}\sin\left(4kx\right) \mathbf{ \hat e}_x, \label{eqn:twoARF}
\end{align}
where ${\bf{F}}_{11}$ and ${\bf{F}}_{12}$ are acoustic radiation forces by SSAWs of frequencies $f$ and $2f$, respectively.
Therefore, the radiation force by W-SSAW can be estimated by adding acoustic radiation forces by two SSAWs when one frequency is twice the other frequency.
\begin{figure}[!t]
	\centering
	\includegraphics[width=\columnwidth]{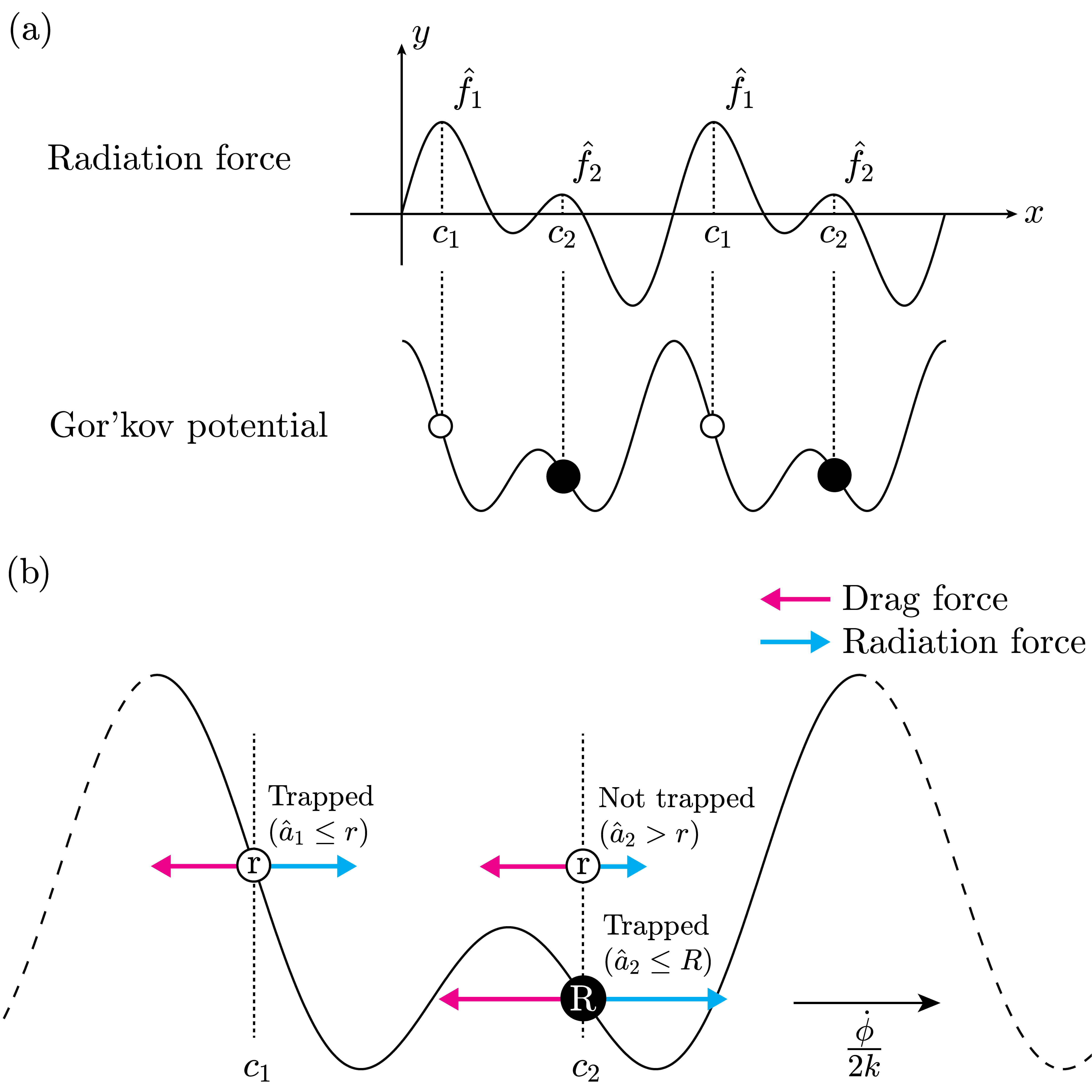}%
	\subfigLabel{fig:W-gorkovs}%
	\subfigLabel{fig:fig1_conditions}%
	\caption{
		\subref{fig:W-gorkovs} Acoustic radiation force of W-SSAW and two equilibrium positions in W-shaped Gor'kov potential under linear phase modulation.
		\subref{fig:fig1_conditions} The conditions for R-particles to be trapped at $c_2$, and r-particles at $c_1$. By lienar phase modulation, Gor'kov potential is moving at the speed of $\dot \phi/2k$. 
		Trapped particles maintain force equilibrium between Stokes' drag and the radiation force.
		\label{fig:fig1}}
\end{figure}
To displace Gor'kov potential of W-SSAW by $\phi/2k$, phase modulations $\phi$ and  $2\phi$ are applied to two SSAWs of frequencies $f$ and $2f$, respectively, since
\begin{equation}
	{\mathbf F}_{1 \rm w}\left(x-\frac{\phi}{2k} \right) = {\mathbf F}_{11}\left( x-\frac{\phi}{2k} \right)+{\mathbf F}_{12}\left(x-\frac{2\phi}{4k}\right).
\end{equation}

For suspended particles subject to W-SSAW with phase modulation, the equation of motion is expressed by
\begin{equation} \label{eqn:eqnmotion}
b \frac{dx}{dt} = f\left(x-\frac{\phi}{2k}\right) V_c,
\end{equation}
where $f\left(\cdot \right)$ denotes acoustic radiation force by W-SSAW per unit volume defined as $F_{1 \rm W}\left(\cdot \right)/V_c$.
Here, the inertia of a particle is neglected since viscous force dominates in microfluidics.
It should be noted that as acoustic radiation force is proportional to the volume of a particle, $f\left(\cdot \right)$ is independent of $V_c$.

For fixed $\phi$, a particle is attracted to minima of Gor'kov potential.
When $\phi$ linearly increases, Gor'kov potential moves at a constant speed of $\dot \phi/2k$.
Acoustic radiation force actuates the particle to follow the minima of moving Gor'kov potential, and the velocity of the particle is $\dot \phi/2k$.
However, as the rate of phase modulation increases, the particle moves faster, and experiences larger drag force.
When the increased drag force exceeds acoustic radiation force, the particle no longer follows the moving Gor'kov potential.
Therefore, there exists the maximum rate of phase change for particles to be trapped, and it is obtained by equating the drag force and the maximum acoustic radiation force.

Suppose $F_{11}$ and $F_{12}$ in \cref{eqn:twoARF} are properly controlled for acoustic radiation force by W-SSAW per unit volume to have two local maxima $\hat f_1$ at $c_1$ and $\hat f_2$ at $c_2$, and $\hat f_1 > \hat f_2$ [\Cref{fig:W-gorkovs}].
From the two local maxima, two maximum rates of phase change will be given.
When drag force and the maximum acoustic radiation force are equal, \cref{eqn:eqnmotion} separates into
\begin{equation} \label{eqn:splited}
b\frac{{dx}}{{dt}} = \hat f_i V_c, \qquad x-\frac{\phi}{2k} = c_i,
\end{equation}
where $i=1,2$.
From \cref{eqn:splited}, the phase to be modulated is given as
\begin{equation} \label{eqn:phimax}
	\phi = \frac{2 k \hat f_i V_c }{b}t+2 k x_0 -2 k c_i,
\end{equation}
where $x_0$ is the initial position of the particle.
Hence, for spherical particles, the maximum rate of phase change $\dot \Phi$ for a particle of radius $a$ to be trapped at equilibrium position $c_i$ is
\begin{equation} \label{eqn:maxphidot}
  \dot \Phi_{i,a} = \frac{2 k \hat f_i V_c}{b} = \frac{4 k \hat f_i}{9 \eta} a^2.
\end{equation}
Rewriting \cref{eqn:maxphidot} with respect to the radius gives
\begin{equation} \label{eqn:minradius}
	\hat a_i \leq a, \qquad \hat a_i = \sqrt {\frac{{9\eta }}{{4k{\hat f_i}}}\dot \phi },
\end{equation}
where $\dot \phi$ denote the given rate of phase change, and $\hat a_i$ is defined to be the critical radius at $c_i$. 
It should be noted that from $\hat f_1 > \hat f_2$, $\hat a_1 < \hat a_2$.

\begin{figure}[!t]
	\centering
	\includegraphics[width=\columnwidth]{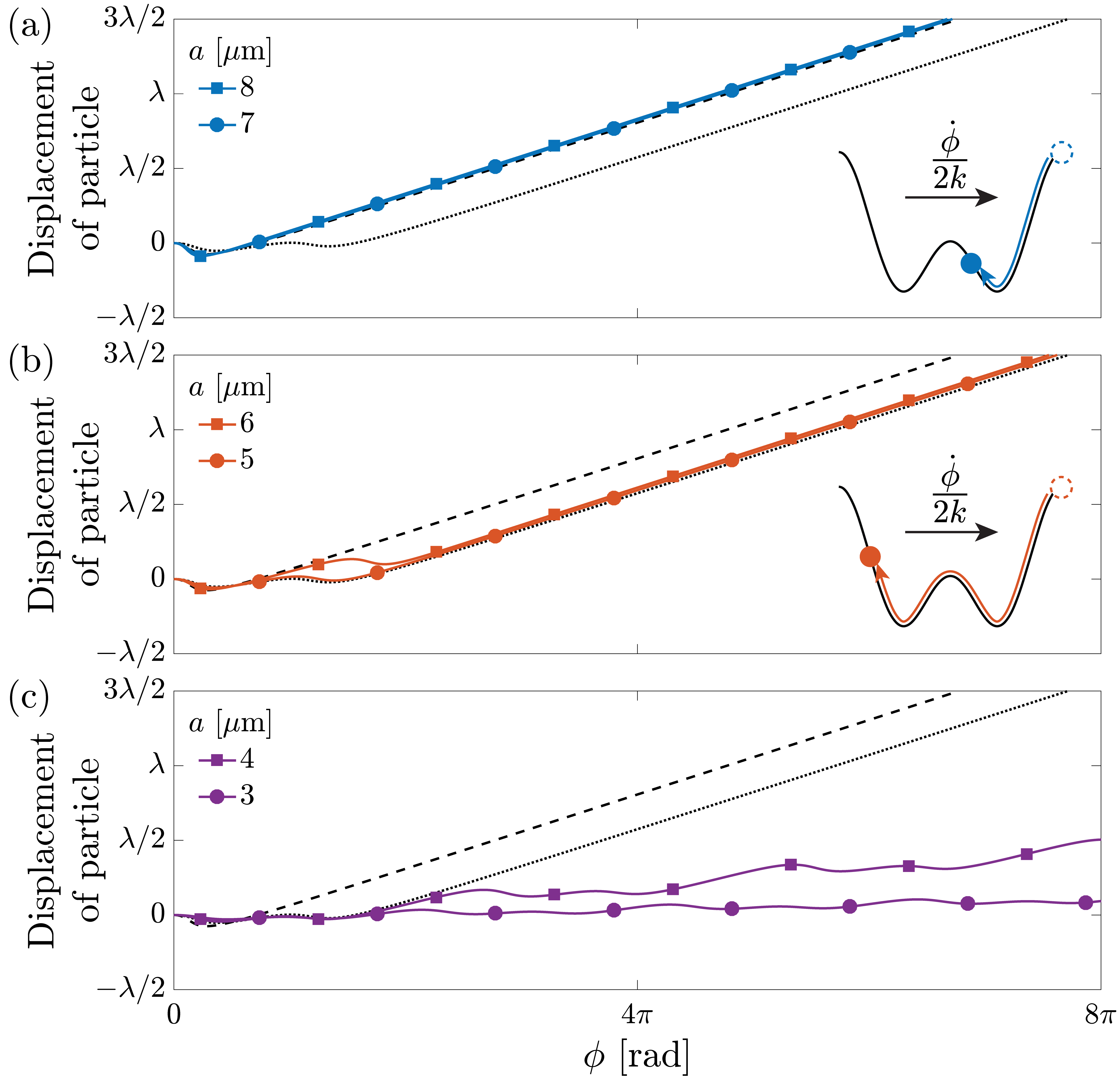}%
	\subfigLabel{fig:fig1_matlab1}%
	\subfigLabel{fig:fig1_matlab2}%
	\subfigLabel{fig:fig1_matlab3}%
	\caption{
		\textbf{Displacements of particles with varius radii under linear phase modulation.}
		Dotted and dashed black lines represent displacements of particles of the critical radii $\hat a_1$ and $\hat a_2$, respectively.
		The \emph{inset} shows the trajectories of particles to the trapped positions on Gor'kov potential.
		\subref{fig:fig1_matlab1} Particles satisfying \cref{eqn:cond1} are trapped at $c_2$.
		\subref{fig:fig1_matlab2} Particles satisfying \cref{eqn:cond2} are trapped at $c_1$.
		\subref{fig:fig1_matlab3} The remaining particles are not captured by W-SSAW exhibiting oscillation.
	}
\end{figure}
As $\hat a_1 \neq \hat a_2$, the groups of particles to be trapped at $c_1$ and $c_2$ are different for a given rate of phase modulation.
Suppose two groups of particles denoted by R-particles and r-particles are of radii $R$ and $r$  $\left(R>r\right)$, respectively.
There exist two possible arrangements into which two groups of particles are separated and patterned: 
{\setlist[enumerate]{noitemsep}
\begin{enumerate}[label=(\roman*)]
	\item R-particles at $c_2$, and r-particles at $c_1$.
	\item R-particles at $c_1$, and r-particles at $c_2$.
\end{enumerate}
}\noindent
Yet, the latter configuration is impossible.
In the latter, r-particles are also trapped at $c_1$ as $\hat a_2 \leq r$ implies $\hat a_1 < r$.
Moreover, R-particles can also be trapped at $c_2$ as $\hat a_2 \leq r$ satisfies $\hat a_2 < R$.
Consequently, both R-particles and r-particles are able to be trapped in  both equilibrium positions eliminating the asymmetry.
Hence, we only consider the former arrangement, which requires the following conditions [\Cref{fig:fig1_conditions}].
\begin{subequations}\label{eqn:ineqphidot}
\begin{enumerate}[label=(\roman*)]
	\item R-particles must be trapped by the smaller maximum of the acoustic radiation force $\hat f_2$ to stay at $c_2$. From \cref{eqn:maxphidot,eqn:minradius},
	\begin{equation} \label{eqn:cond1}
		\dot \phi \leq \dot \Phi_{2,R},\qquad \hat a_2 \leq R.
	\end{equation}
	\item \label{cond:cond3} r-particles must be trapped by the larger maximum of the acoustic radiation force $\hat f_1$ to be trapped at $c_1$. From \cref{eqn:maxphidot,eqn:minradius},
	\begin{equation} \label{eqn:cond2}
		 \dot \phi \leq \dot \Phi_{1,r},\qquad \hat a_1 \leq r.
	\end{equation}
	\item r-particles must not be trapped by the smaller maximum of the acoustic radiation force $\hat f_2$ to be passed to $c_1$. From the negation of \cref{eqn:maxphidot,eqn:minradius},
	\begin{equation} \label{eqn:cond3}
		\dot \phi > \dot \Phi_{2,r},\qquad \hat a_2 > r.
	\end{equation}    
\end{enumerate}
\end{subequations}

Numerical analysis is performed with the material properties listed in \cref{table:table1} to show that two groups of particles are selectively trapped at each equilibrium position. The displacements of particles of various diameters are examined at the arbitrarily chosen rate of phase change.
The result reveals that R-particles satisfying \cref{eqn:cond1} are trapped at $c_2$ [\Cref{fig:fig1_matlab1}], and r-particles satisfying \cref{eqn:cond2} at $c_1$ [\Cref{fig:fig1_matlab2}].
The remaining particles are not trapped by any equilibrium positions exhibiting oscillatory motion [\Cref{fig:fig1_matlab3}].

\begin{figure}[!t]
	\includegraphics[width=\columnwidth]{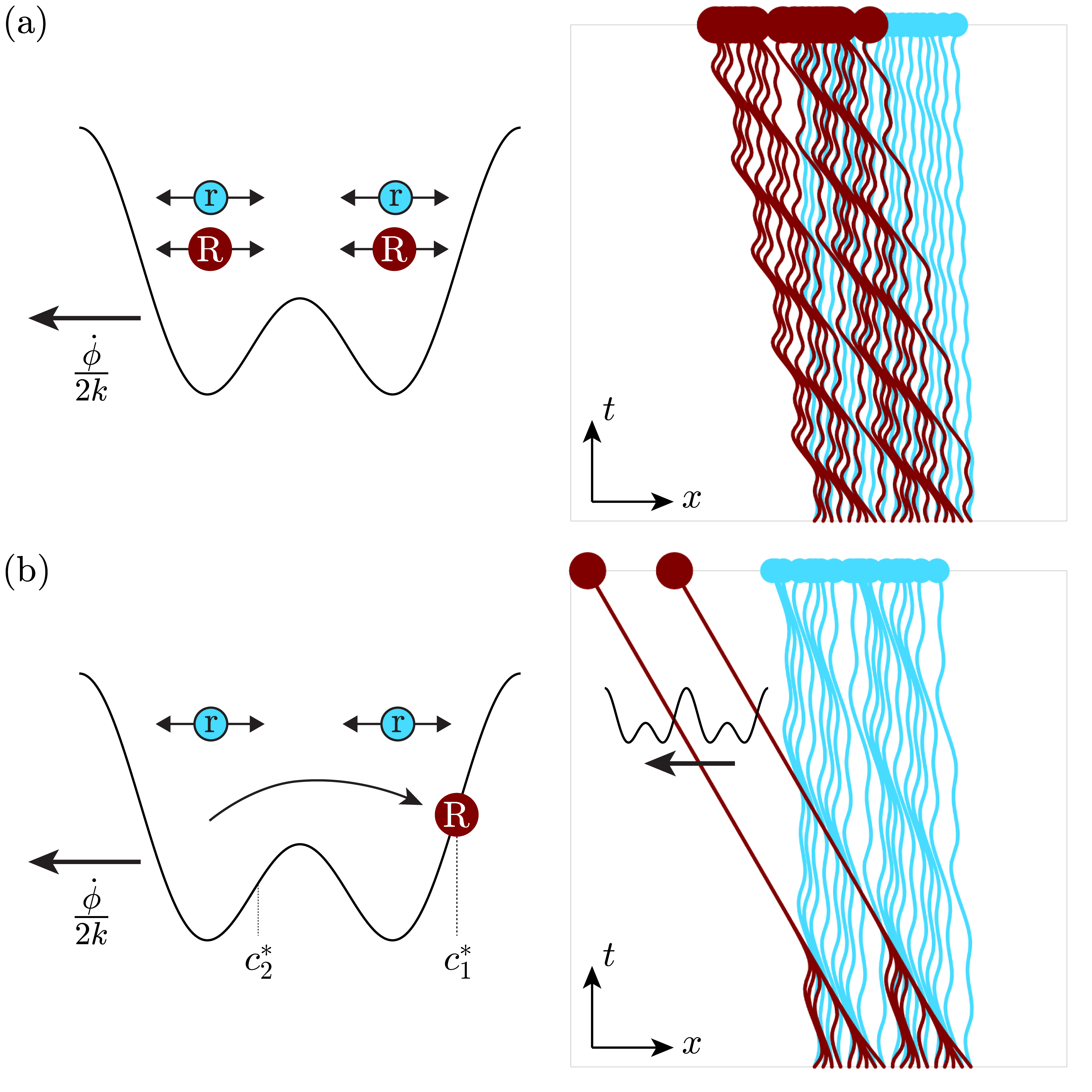}%
	\subfigLabel{fig:fig2_state1}%
	\subfigLabel{fig:fig2_state2}%
	\caption{
		\textbf{Motions and displacements of particles during the preprocess.}
		Red and blue spheres represent R- and r- particles, respectively.
		\subref{fig:fig2_state1} The rate of phase change $\dot \phi$ is so high that both particles are not trapped by any equilibrium position but oscillate.
		\subref{fig:fig2_state2} $\dot \phi$ is properly selected to trap and aggregate R-particles at $c_1^\ast$.
		The trajectories of r-particles remain dispersed due to oscillatory motion induced by high rate of phase change.
		$c_1^\ast$ and $c_2^\ast$ denote the corresponding positions of $c_1$ and $c_2$ when Gor'kov potential translates in the opposite direction, respectively.
		\label{fig:fig2}}
\end{figure}
\begin{figure*}[!t]
	\begin{minipage}[l]{0.667\textwidth}%
		\includegraphics[width=\textwidth]{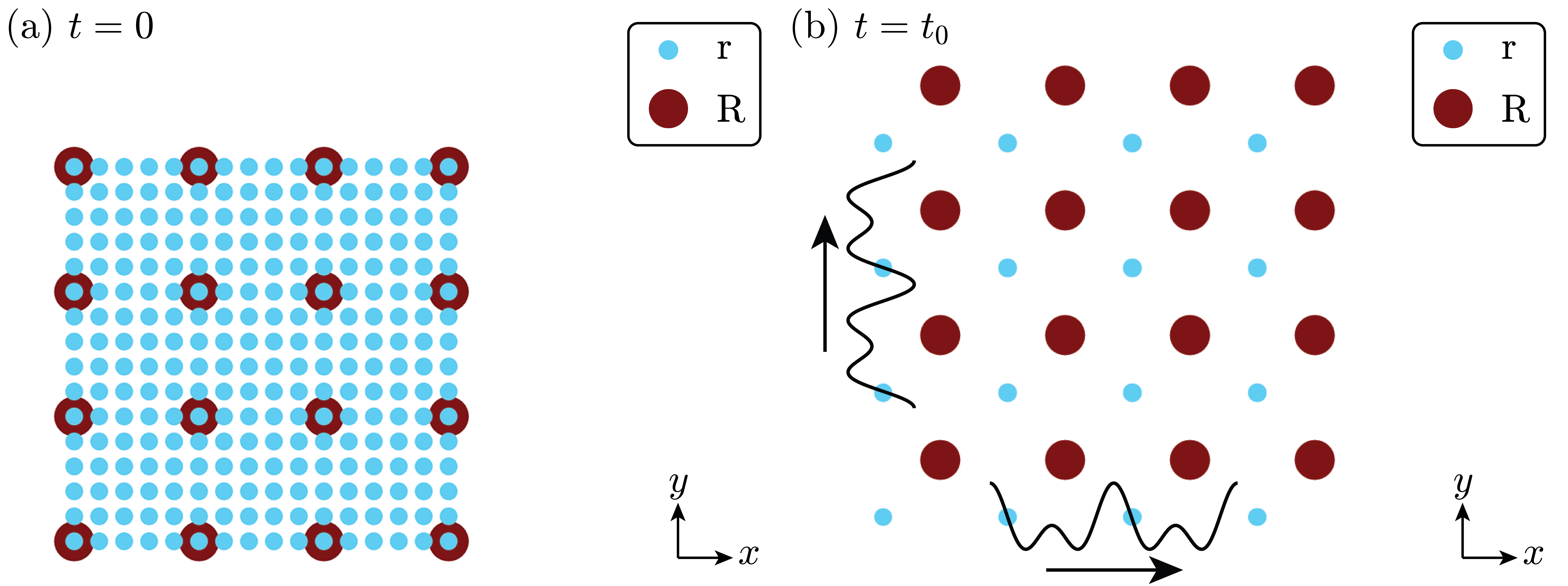}%
	\end{minipage}\hfill%
	\subfigLabel{fig:fig3_COMSOL-Before}%
	\subfigLabel{fig:fig3_COMSOL-After}%
	\begin{minipage}[c]{0.3\textwidth}%
		\caption{
			\textbf{Results of the preprocess and two-dimensional alternate patterning.}
			\subref{fig:fig3_COMSOL-Before} Initial arrangement of particles as a result of the preprocess using two-dimensional W-SSAW at the properly high rate of phase modulation. R-particles are patterned at $c_1^\ast$, whereas r-particles remain dispersed.
			\subref{fig:fig3_COMSOL-After} Two dimenional alternate patterning by X- and Y-directional W-SSAWs. R-particles are trapped at $c_2$, and r-particles at $c_1$ forming alternate grid patterns.
			\label{fig:fig3}}
	\end{minipage}%
\end{figure*}

Since $\hat a_{1} < R$, from \cref{eqn:cond2}, R-particles are also able to be trapped at $c_1$.
In other words, if R-particles are initially placed in the interval $I= \left[c_1,c_2\right)$, both R-particles and r-particles will be trapped at $c_1$.
This problem can be tackled by the preprocess which positions R-particles in the outside of $I$.
Though this preprocess can be achieved by conventional SSAW-based patterning methods, it is unfavorable in that SSAW aggregates not only R-particles but also r-particles into pressure nodes or antinodes.
Once particles are aggregated, interparticle force arising from scattering of neighboring particles becomes dominant to invalidate \cref{eqn:eqnmotion}.\cite{laurell2007chip}
In addition, the aggregation of particles has the effect of the aggregated particles having larger radius apparently.
Considering W-SSAW separates particles depending on size difference, if r-particles aggregate during the preprocess, separating out r-particles to $c_1$ becomes difficult.
Hence, the preprocess should not aggregate r-particles.

Patterning R-particles without aggregating r-particles can be accomplished using W-SSAW at a properly high rate of phase modulation.
When the rate of phase change is too high for both R- and r-particles to follow the moving Gor'kov potential, \emph{i.e.}
\begin{equation}
\dot \phi > \max\left\{\dot \Phi_{1,R}, \dot \Phi_{1,r}, \dot \Phi_{2,R}, \dot \Phi_{2,r} \right\},
\end{equation}
both particles oscillates [\Cref{fig:fig2_state1}].
At gradually decreasing $\dot \phi$, one can find the rate of phase change satisfying
\begin{equation}
\dot \Phi_{1,R} \geq \dot \phi > \max\left\{\dot \Phi_{1,r}, \dot \Phi_{2,R}, \dot \Phi_{2,r} \right\}.
\end{equation}
At that rate, only R-particles are trapped at $c_1^\ast$, the corresponding equilibrium position of $c_1$ in the opposite direction, which is in the outside of $I$ [\Cref{fig:fig2_state2}].
It should be noted that r-particles stay dispersed.
Hence, W-SSAW combined with the properly high rate of phase modulation can pattern R-particles without aggregation of r-particles.

\begin{figure*}[!t]
	\begin{minipage}[l]{0.333\textwidth}%
		\includegraphics[width=2\textwidth]{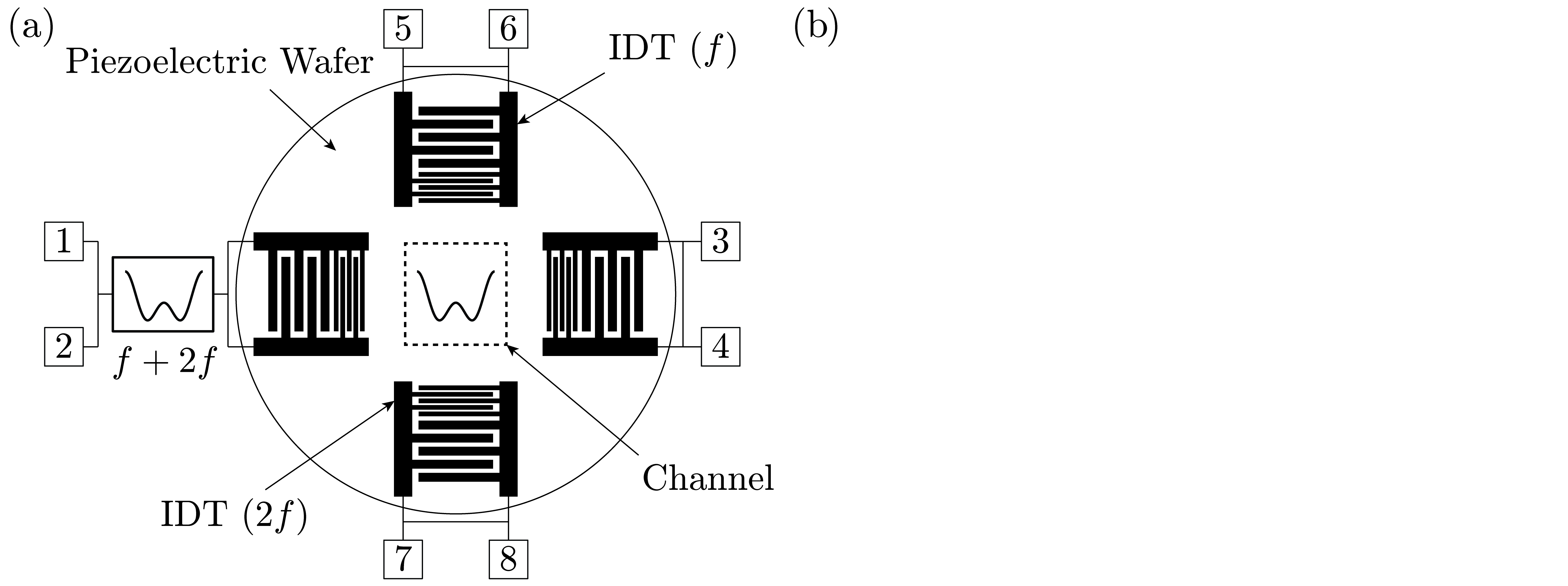}%
	\end{minipage}%
	\begin{minipage}[l]{0.333\textwidth}%
		{	\renewcommand{\arraystretch}{1.125}%
			\setlength\tabcolsep{0.8em}%
			\centering%
			\begin{tabular}{ccc}
				\hhline{---} 
				& & \vspace{-1em} \\
				\textbf{Port}  & \textbf{Frequency} & \textbf{Phase} \\ \hhline{===}
				1 & $f+\Delta f/2$ & $\phi_{11}/2$ \\
				2 & $2f+2\Delta f/2$ & $\phi_{12}/2$ \\
				3 & $f-\Delta f/2$ & $\phi_{11}/2$ \\			
				4 & $2f-2\Delta f/2$ & $\phi_{12}/2$ \\
				5 & $f+\Delta f/2$ & $\phi_{21}/2$ \\
				6 & $2f+2\Delta f/2$ & $\phi_{22}/2$ \\
				7 & $f-\Delta f/2$ & $\phi_{21}/2$ \\			
				8 & $2f-2\Delta f/2$ & $\phi_{22}/2$ \\
				\hhline{---}
			\end{tabular}}%
		\end{minipage} \hfill%
		\subfigLabel{fig:fig4_expSetup}%
		\subfigLabel{fig:channel}%
		\begin{minipage}[c]{0.3\textwidth}
			\caption{
				\subref{fig:fig4_expSetup} Experimental setup  for two-dimensional alternate patterning using W-SSAW.
				\subref{fig:channel} Configuration of frequency and phase of each port of a function generator.
				The shifted freqeuencies are to simulate linear phase modulation.
				The phase differences between X- and Y-directional W-SSAWs are maintained as \SI{90}{\degree}.
				\label{fig:fig4}}
		\end{minipage}
	\end{figure*}
One-dimensional W-SSAW can be extended to two-dimensional alternate patterning.
Similar to \cref{eqn:pv1w},
pressure field ${p}_{{\rm{in}}}$ and velocity vector field ${{\bf{v}}_{{\rm{in}}}}$ of two-dimensional W-SSAW are superposed as
\begin{subequations}
\begin{align}
{p}_{{\rm{in}}}  & =  {p}_{1{\rm{w}}}\left(x,t\right) +  {p}_{2{\rm{w}}} \left(y,t\right)\\
				& = p_{11} + p_{12} + p_{21} + p_{22}, \nonumber \\
	{{\bf{v}}_{{\rm{in}}}}  & =  {{\bf{v}}_{1{\rm{w}}}}\left(x,t\right) +  {{\bf{v}}_{2{\rm{w}}}}\left(y,t\right) \\
				& = {\mathbf v}_{11} + {\mathbf v}_{12} + {\mathbf v}_{21} + {\mathbf v}_{22}, \nonumber
\end{align}
\end{subequations}
where ${p}_{2{\rm{w}}}$ and ${{\bf{v}}_{2{\rm{w}}}}$ are pressure field and velocity vector field of Y-directional W-SSAW.
The root mean square for $p_{\rm in}$ is expressed as
\begin{widetext}	
\begin{align}
\left\langle {p_{{\rm{in}}}^2} \right\rangle & ={\left\langle {p_{11}^2} \right\rangle  + \left\langle {p_{12}^2} \right\rangle  + \left\langle {p_{21}^2} \right\rangle  + \left\langle {p_{22}^2} \right\rangle } +2\Big({ \left\langle {{p_{11}}{p_{12}}} \right\rangle  + \left\langle {{p_{21}}{p_{22}}} \right\rangle  + \left\langle {{p_{11}}{p_{22}}} \right\rangle  + \left\langle {{p_{12}}{p_{21}}} \right\rangle }\Big) +2 \Big({\left\langle {{p_{11}}{p_{21}}} \right\rangle  + \left\langle {{p_{12}}{p_{22}}} \right\rangle }\Big) \label{eqn:orthogonalPTerms} \\
	& = {\left\langle {p_{11}^2} \right\rangle  + \left\langle {p_{12}^2} \right\rangle  + \left\langle {p_{21}^2} \right\rangle  + \left\langle {p_{22}^2} \right\rangle +2 \Big({\left\langle {{p_{11}}{p_{21}}} \right\rangle  + \left\langle {{p_{12}}{p_{22}}} \right\rangle }\Big)}. 
\end{align}
The terms in the first parenthesis in \cref{eqn:orthogonalPTerms} vanishes due to the orthogonality of trigonometric functions as in \cref{eqn:rmsp1w}.
On the other hand,
\begin{align}
\left\langle {{p_{11}}{p_{21}}} \right\rangle & =  {A_{11}}{A_{21}}\cos \left( {kx} \right)\cos \left( {ky} \right)\left\langle {\sin \left( {\omega t + {\phi _{11}}} \right)\sin \left( {\omega t + {\phi _{21}}} \right)} \right\rangle \nonumber \\
	& =  {A_{11}}{A_{21}}\cos \left( {kx} \right)\cos \left( {ky} \right) \cdot \frac{1}{2}{\cos \left( {{\phi _{11}} - {\phi _{21}}} \right)}. \label{eqn:p1121}
\end{align}
\end{widetext}
When $\phi_{11}-\phi_{21}=\pm \pi/2$, \cref{eqn:p1121} vanishes. Similarly, choosing $\phi_{12}-\phi_{22}=\pm \pi/2$, we have $\left\langle {{p_{12}}{p_{22}}} \right\rangle =0$.
Hence,
\begin{equation}
\left\langle p_{\rm in}^2 \right\rangle = \left\langle p_{1 \rm w}^2 \right\rangle  + \left\langle p_{2 \rm w}^2 \right\rangle.
\end{equation}
On the other hand, using $\left\langle {{{\bf{v}}_{11}} \cdot {{\bf{v}}_{21}}} \right\rangle  = \left\langle {{{\bf{v}}_{12}} \cdot {{\bf{v}}_{22}}} \right\rangle  = 0$,
\begin{equation}
\left\langle {{\bf{v}}_{{\rm{in}}}^2} \right\rangle  = {\rm{ }}\left\langle {{\bf{v}}_{1{\rm{w}}}^2} \right\rangle  + \left\langle {{\bf{v}}_{2{\rm{w}}}^2} \right\rangle.
\end{equation}
From \cref{eqn:gorkov}, Gor'kov potential of two-dimensional W-SSAW denoted by $U\left(x,y\right)$ is expressed as
\begin{equation}
	U\left(x,y\right) = U_{1 \rm w}\left(x\right) + U_{2 \rm w}\left(y\right),
\end{equation}
where $U_{2 \rm w}$ denotes Gor'kov potential of Y-directional W-SSAW.
Therefore, acoustic radiation force for two-dimensional W-SSAW denoted by ${\mathbf F}\left(x,y\right)$ is given by
\begin{equation} \label{eqn:2d_ARF}
	{\mathbf{F}}\left(x,y\right) =  {\mathbf F}_{1\rm w}\left(x\right) + {\mathbf F}_{2 \rm w} \left(y\right),
\end{equation}
where ${\mathbf F}_{2 \rm w}$ is acoustic radiation force for Y-directional W-SSAW.
As \cref{eqn:stokes,eqn:2d_ARF} can be decoupled into X- and Y-directional components, the equation of motion is decoupled.
It is followed that two-dimensional alternate patterning is performed by applying one-dimensional alternate patterning independently in each direction.

Numerical analysis using Particle Tracing Module in COMSOL Multiphysics\textsuperscript{\textregistered} is performed to show two-dimensional alternate patterning.
Suppose the preprocess has already been performed, so that R-particles are initially placed at $c_1^\ast$, and r-particles are randomly distributed [\Cref{fig:fig3_COMSOL-Before}].
The result shows that two-dimensional W-SSAW is capable of patterning particles alternately in two-dimension to create an alternate grid pattern [\Cref{fig:fig3_COMSOL-After}].

In the experiment, two pairs of interdigital transducers (IDT) is prepared to radiate two SSAWs of frequencies $f$ and $2f$ to generate W-SSAW [\Cref{fig:fig4_expSetup}].
As the bandwidth of an IDT is narrow enough that a pair of IDTs designed for a certain frequency only generates SSAW of the designed frequency.
It implies that even if the combined signal of two frequencies is applied, each IDT only radiates its corresponding frequency from the mixed signal.
Linear phase modulation can be simulated by shifting frequency.
For example, on port 1, 
\begin{align}
y & = y_0 \cos (kx-2\pi f t + (\dot \phi/2 ) t + \phi_{11}/2 ) \nonumber \\
  & = y_0 \cos \left(kx-2\pi \left(f+\Delta f /2\right) t + \phi_{11}/2\right),
\end{align}
where $\Delta f=\dot\phi /2\pi$.
Similarly, the frequency and the phase of each port can be determined to have phase differences between X- and Y-directional W-SSAW and to simulate linear phase modulation [\Cref{fig:channel}]. Again, $\phi_{11}-\phi_{21}=\pm \pi/2$ and $\phi_{12}-\phi_{22}=\pm \pi/2$ must be maintained.

The acoustic method of patterning microparticles into an alternate grid pattern using W-SSAW is developed.
From the orthogonality of trigonometric function, W-SSAW constructed by two SSAWs of $f$ and $2f$ is shown to have W-shaped Gor'kov potential.
When linear phase modulation is applied to W-SSAW, there emerge two asymmetric equilibrium positions for particles to be trapped.
By the asymmetry, R- and r-particles are positioned into different equilibrium positions to have an alternate pattern.
Furthermore, two-dimensional patterning is achieved by controlling the phase differences \SI{90}{\degree} in X- and Y-directional W-SSAWs.
Moreover, the potential strength is that W-SSAW can pattern three or more than groups of particles into alternate grid patterns simultaneously.
Specifically, when three SSAWs of frequencies $f$, $2f$, and $3f$ are superposed, similar to two SSAWs, acoustic radiation force by three SSAWs can be estimated by linear addition.
One can, then, find three different local maxima to have three asymmetric conditions for particles to be trapped.
Hence, three or more than groups of particles can be patterned to form complex grid pattern.
It is believed that patterning different groups of particles into an alternate grid pattern using W-SSAW is beneficial for biological applications such as tissue engineering.
\bibliography{WSSAW}

\clearpage
\setcounter{equation}{0}
\renewcommand{\theequation}{S\arabic{equation}}
\setcounter{table}{0}
\renewcommand{\thetable}{S\arabic{table}}
\setcounter{figure}{0}
\renewcommand{\thefigure}{S\arabic{figure}}

\section*{Supplementary Material}

{\renewcommand{\arraystretch}{1.125}
	\begin{table}[!h]
		\centering
		\begin{tabularx}{.5\textwidth}{Xcrl}
			\hline & & & \vspace{-1em}\\
			\textbf{Material Properties}  & & & \\ \hhline{====}
			Water  & & & \\
			\hline
			Viscosity & $\eta$ & 0.98 & \si{\milli\pascal\second} \\
			Compressibility & $\beta_w$ & 448 & \si{\per\tera\pascal} \\
			Density & $\rho_w$ & 998 & \si{\kilogram\per\cubic\meter} \\
			\hline
			Polystyrene Particles & & & \\
			\hline
			Compressibility & $\beta_c$ & 249 & \si{\per\tera\pascal} \\
			Density & $\rho_c$ & 1050 & \si{\kilogram\per\cubic\meter} \\
			Diameter of R-Particle & $2R$ & 15 & \si{\micro\meter} \\
			Diameter of r-Particle & $2r$ & 10 & \si{\micro\meter} \\
			\hline
			Acoustic Waves & & & \\
			\hline
			Acoustic Pressure & $p_0$ & 100 & \si{\kilo\pascal} \\
			Wavelength & $\lambda$ & 280 & \si{\micro\meter} \\
			\hline
			W-SSAW & & & \\
			\hline
			Rate of Phase Change & $\dot \phi$ & 8 & \si{\radian\per\second}\\
			First Critical Radius$^\dagger$ & $\hat a_1$ & 9.20 & \si{\micro\meter} \\
			Second Critical Raidus$^\dagger$ & $\hat a_2$ & 13.2 &  \si{\micro\meter} \vspace{1em} \\
			\multicolumn{4}{l}{$^\dagger$ Calculated from \cref{eqn:minradius}} \\
			\hline
		\end{tabularx}
		\captionsetup{width=.5\textwidth}
		\caption{Properties used for the numerical analysis.\label{table:table1}}
	\end{table}}
\end{document}